\documentclass[floatfix,reprint,pra,aps,showpacs]{revtex4-1}
\usepackage{graphicx}
\usepackage[normalem]{ulem}

\begin{document}

\title{The structure of a water monolayer on the anatase TiO$_{\mathbf{2}}$ (101) surface}

\author{Christopher E. Patrick}
\author{Feliciano Giustino}
\email{feliciano.giustino@materials.ox.ac.uk}
\affiliation{Department of Materials, University of Oxford, Parks Road,
Oxford OX1 3PH, United Kingdom}

\begin{abstract}
Titanium dioxide (TiO$_2$) plays a central role in the study of artificial photosynthesis,
owing to its ability to perform photocatalytic water splitting. Despite over four decades 
of intense research efforts in this area, there is still some debate over the nature 
of the first water monolayer on the technologically-relevant anatase TiO$_2$ (101) surface. 
In this work we use first-principles calculations to reverse-engineer the experimental
high-resolution X-ray photoelectron spectra measured
for this surface in [Walle {\it et al.}, J.\ Phys.\ Chem.\ C 115, 9545 (2011)],
and find evidence supporting the existence of a mix of dissociated and molecular 
water in the first monolayer.
Using both semilocal and hybrid functional calculations
we revise the current understanding of the adsorption energetics by showing that 
the energetic cost of water dissociation is reduced 
via the formation of a hydrogen-bonded hydroxyl-water complex. 
We also show that 
such a complex can provide an explanation of an unusual superstructure 
observed in high-resolution 
scanning tunneling microscopy experiments. 
\end{abstract}

\date{\today}
\maketitle

\section{Introduction}

The realization of artificial photosynthesis hinges on our ability to
engineer materials and devices which can effectively convert 
solar energy into fuels~\cite{Tachibana2012}.  
Since the discovery of photocatalytic water splitting more than four decades ago~\cite{Fujishima1972} 
titanium dioxide (TiO$_2$)
has been a lead contender in this field, and has found applications in a variety of cutting-edge research 
areas~\cite{Gratzel20012,Paz1995,Roy2010}.
As water is ubiquitous in all these applications it is not surprising that numerous studies 
have been devoted to understanding the physics of the interface between TiO$_2$ and 
 H$_2$O, particularly the rutile (110) 
surface~\cite{Kristoffersen2013,Liu2010,Allegretti2005,Matthiesen2009,Brookes2001,Lindan2005}.
However in the context of nanotechnology the anatase (101) surface is even more interesting,
since this is the most common surface in nanostructured 
TiO$_2$~\cite{Diebold2003,Shklover1998,Lazzeri2001}.

A fundamental question in this area is whether H$_2$O molecules 
adsorbed on the anatase surface are intact (molecular adsorption), or
dissociated into a hydroxyl group and a proton (dissociative adsorption).
Until very recently 
experimental and theoretical studies consistently promoted the notion that only molecular 
adsorption occurs on the pristine anatase (101) surface.
In fact, the density-functional theory (DFT) study of 
Ref.~\cite{Vittadini1998} found molecular adsorption to be energetically favorable 
over dissociation by 0.44~eV/0.28~eV per H$_2$O molecule (at low/monolayer coverage, respectively).
A few years later the authors of Ref.~\cite{Herman2003} concluded in favor of the molecular
adsorption scenario, based on their analysis of  X-ray photoelectron spectroscopy (XPS) 
and temperature-programmed 
desorption (TPD) experiments.
Furthermore, apart from a couple of exceptions~\cite{Raju2013,Arrouvel2004}, most computational studies
continue to find molecular adsorption to be energetically favorable~\cite{Sun2010}.

More recently, scanning tunneling microscopy (STM) experiments 
combined with DFT calculations identified molecular water on the anatase surface at very 
low coverage~\cite{He2009}. The interpretation of STM images for higher coverages 
proved more challenging due to the appearance of an unexpected 2$\times$2 superstructure.
Then, in contrast with the prevailing view of molecular adsorption of water on anatase, 
the authors of a recent high-resolution XPS study~\cite{Walle2011} reported the presence 
of OH groups at the surface, and challenged the  accepted model of the 
interface. From an analysis of the XPS spectral intensities the authors of 
Ref.~\cite{Walle2011} proposed that, at monolayer coverage, 23$\pm$5\% of the water 
molecules dissociate into a hydroxyl and a proton. 

In this work we test the proposal of dissociative water adsorption of Ref.~\cite{Walle2011} 
using first-principles calculations. Our approach consists of calculating the O-$1s$ core-level 
spectra of interface models where water is adsorbed either molecularly or dissociatively, 
and comparing our {\it ab initio} spectra to the experimental data of Ref.~\cite{Walle2011}.
The present analysis is motivated by the very high accuracy ($<$0.2~eV) that can now be achieved 
in DFT calculations of core-electron binding energies, as demonstrated by recent work~\cite{Patrick2011,
Pozzo2011,Binder2012}.

As we show below, the comparison between our calculated spectra and experiment
provides evidence for the presence of dissociated water 
in the experiments of Ref.~\cite{Walle2011} 
on anatase TiO$_2$ (101) at monolayer coverage.
This finding prompted us to re-examine the factual evidence 
underlying the molecular model of the TiO$_2$:H$_2$O 
interface~\cite{Vittadini1998,Sun2010,Herman2003,He2009},
namely the unfavorable energetics calculated for dissociative adsorption and the measured STM images.
We found that the energetics of H$_2$O dissociation on anatase is
altered when two water molecules are adsorbed on adjacent sites. In these conditions 
the energetic cost of dissociation is compensated by the energy gain due to the formation 
of the H-bonded complex HO$\cdot\cdot\cdot$H$_2$O, and dissociative  adsorption becomes possible.
This H-bonded complex also leads to a 2$\times$2 superstructure like that observed in
the STM images of Ref.~\cite{He2009}.

\section{Methodology}

Calculations were performed within the generalized-gradient approximation 
to DFT of 
Ref.~\cite{Perdew1996}
(the PBE functional), employing pseudopotentials, plane-wave basis sets and periodic boundary 
conditions~\cite{Vanderbilt1990, Lazzeri2001,Patrick2011,Binder2012,Makov1995,Ackland1997} 
as implemented in the \texttt{Quantum ESPRESSO} distribution~\cite{quantumespresso} 
(technical details are given in Appendix~\ref{app.dft}).
For the sake of accuracy, adsorption energy calculations were also performed 
with the PBE0 hybrid functional~\cite{Adamo1999}.
Core-level shifts were calculated as differences in DFT total energies between systems containing 
a core-hole, as obtained by replacing the pseudopotential of a given oxygen atom by one containing
a hole in the 1$s$ level~\cite{Pehlke1993,Pasquarello1996}. 
In order to minimize the unphysical interaction of charged replicas~\cite{Binder2012,Makov1995}
we used model systems containing up to 432 atoms and 3200 electrons.
As in experiment we can determine 
core-electron binding energies only modulo a constant, which here we choose such that the 
calculated bulk O-1$s$ peak is aligned to the experimental spectra. 

The photoelectron spectrum 
$I(E)$ was determined using the expression $I(E) = \sum_i \exp(-z_i/\lambda) \delta(E-E_i)$. 
Here $E_i$ is the O-$1s$ binding energy of the atom $i$ at a distance $z_i$ below the surface, 
$\delta$ is the Dirac delta function, and the exponential accounts for the finite escape depth 
$\lambda$ of the photoelectrons~\cite{Hufnerbook2}. In the sum we explicitly included 
the contributions of the adsorbates and those of the 6 topmost TiO$_2$ layers of the slab. 
In order to also describe the contribution of bulk TiO$_2$ far from the surface, we added 
the contributions from an infinite number of ``virtual'' atoms, whose positions were determined 
by the bulk anatase lattice. The O-1$s$ binding energy of virtual atoms was set to the average 
value taken over the central 4 TiO$_2$ layers of our slab model. The photoelectron escape depth 
was set to $\lambda=$3.5~\AA, consistent with the range 2--12~\AA\ determined from the universal 
curve~\cite{Hufnerbook2} for kinetic energies of $\sim$80~eV~\cite{Walle2011}. 
Different values of $\lambda$ change the relative strengths of the bulk and adsorbate
contributions to the spectrum, but preserve the adsorbate features~(see Appendix~\ref{app.lambda}).
The Dirac delta 
functions were replaced by gaussians in order to account for vibrational broadening, core-hole 
lifetimes, and instrumental resolution. We evaluated the vibrational broadening using 
importance-sampling Monte Carlo calculations as in Ref.~\cite{Patrick2013}. Quantum zero-point 
motion was found to induce a significant broadening, with a FWHM of $\sim$1~eV. Accordingly 
we set the FWHM in the gaussian broadening to 1.2~eV. This choice should be representative of 
all the aforementioned broadening effects. For the purpose of resolving individual contributions, 
in the following figures we also show spectra obtained using a very small artificial 
broadening of 0.1~eV. In all the figures a linear background was subtracted from the experimental data.

\section{Results and discussion}

\begin{figure}[t!]
\includegraphics{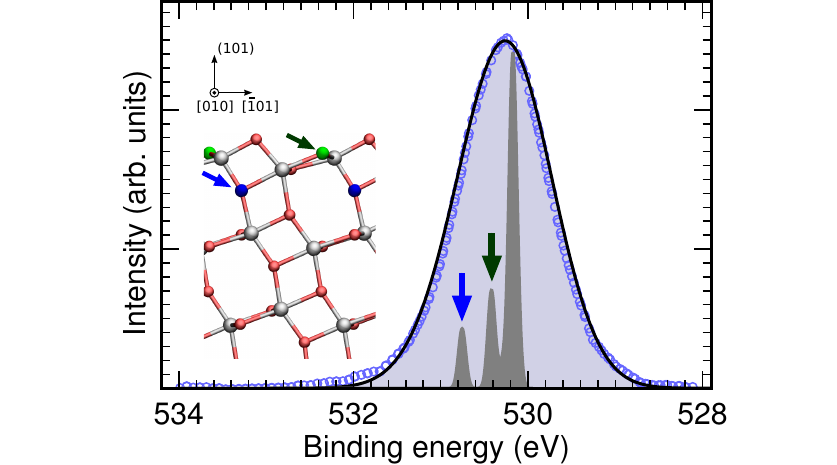}
\caption{
\label{fig.1} (Color online)
Calculated O-$1s$ core-level spectrum of the clean anatase TiO$_2$ (101) surface, compared 
to experiment~\cite{Walle2011} (blue circles). The black line and dark shaded area correspond 
to a gaussian broadening of 1.2~and~0.1~eV, respectively.
The inset shows a ball-and-stick model of the anatase (101) surface, with O atoms in red 
and Ti atoms in white. The O atoms represented in green and blue give rise to the two 
satellites indicated by the arrows.
}
\end{figure}

Figure~\ref{fig.1} shows a ball-and-stick representation of the clean anatase TiO$_2$ (101) surface
and its corresponding O-1$s$ core-level spectrum.
Most of the oxygen atoms in the
substrate, including the twofold coordinated O atoms at the surface, contribute to a single peak 
at 530.17~eV, with a very small spread of 0.05~eV (right peak of the gray shaded area).
Two important exceptions appear as satellites at 530.42~and~530.76~eV, indicated by
arrows. These satellites arise from the structural relaxation taking place at the surface~\cite{Lazzeri2001}.
The fivefold coordinated 
Ti atoms at the surface relax towards the substrate, causing a 0.1~\AA\ contraction of the 
Ti--O bond formed with a nearest neighbor O atom underneath (indicated in blue). This contraction 
is responsible for an increase in the O-1$s$ binding energy by 0.59~eV with respect to bulk (blue arrow).
Similarly the surface relaxation 
also causes a distortion in the second O layer (marked in green), raising these atoms by 0.34~\AA \ 
out of their ideal planar bonding with the 3 Ti neighbors. This distortion is responsible 
for an increase in binding energy by 0.25~eV (green arrow). Figure~\ref{fig.1} shows that 
our calculations are in very good agreement with the experimental spectrum reported 
in Ref.~\cite{Walle2011} for the clean anatase surface (light blue circles),
with the structural relaxation effects accounting for the 
slight asymmetry in the experimental lineshape.

\begin{figure}[t!]
\includegraphics{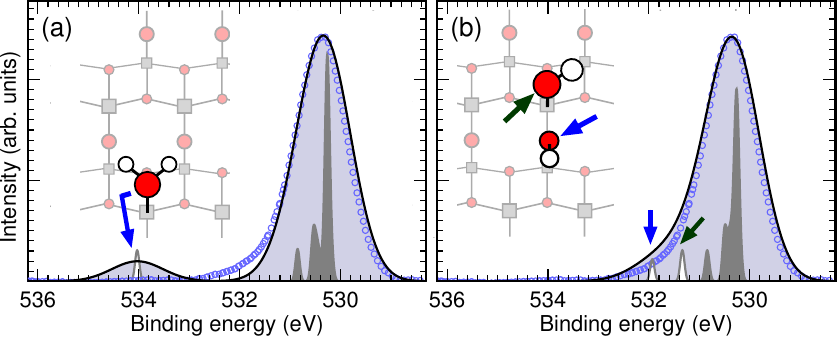}
\caption{
\label{fig.2}  (Color online)
Calculated O-$1s$ core-level spectra for the TiO$_2$:H$_2$O interface at low coverage for
molecular (a) and dissociative (b) adsorption, compared with the experimental 
O-$1s$ spectrum at 0.1~ML coverage (blue circles, spectrum at 300~K in 
Fig.~3 of Ref.~\cite{Walle2011}). 
Black and gray lines correspond to a gaussian broadening of 1.2~and~0.1~eV, 
respectively. The insets show ball-and-stick models viewed from above with
squares and circles representing Ti and O.
The arrows identify O-$1s$ shifts of particular atoms. Dark shaded 
areas indicate the contributions from the TiO$_2$ substrate. 
}
\end{figure}

Next we investigate the core-level spectra in the presence of water at low coverage.
We consider both molecular and dissociative adsorption, as shown by the atomistic models 
in Fig.~\ref{fig.2}. In molecular adsorption the H$_2$O molecule forms a dative bond with 
a fivefold Ti atom at the surface [bond length 2.32~\AA, Fig.~\ref{fig.2}(a)].
In dissociative adsorption the OH group bonds covalently 
to a fivefold Ti atom (bond length 1.82~\AA), and the proton bonds to a twofold O atom,
thereby yielding an additional hydroxyl group [Fig.~\ref{fig.2}(b)]. 
We explored models differing in the separation between the two OH groups (Appendix~\ref{app.OH})
but here we focus on the model
shown in Fig.~\ref{fig.2}(b) previously described as 
the ``intrapair configuration''~\cite{Vittadini1998} or ``pseudo-dissociated
water''~\cite{Walle2011}.

Considering molecular adsorption first [Fig.~\ref{fig.2}(a)], the calculated spectrum exhibits 
a new feature at 534.03~eV arising from the O atom of the water molecule. At the same time 
in the energy range 528--532~eV the spectrum is identical to that of the clean 
surface (Fig.~\ref{fig.1}). This observation is explained by noting that the 
dative H$_2$O--Ti bond does not induce any significant distortion in the underlying substrate.
In the case of dissociative adsorption
the spectrum exhibits two new contributions at 531.34/531.93~eV,
arising from the inequivalent OH groups marked by the green and blue arrows. 
Other configurations (Appendix~\ref{app.OH}) yield essentially identical results.

In Figs.~\ref{fig.2}(a)~and~\ref{fig.2}(b) we compare our calculated spectra at low coverage (0.25~ML) 
with the XPS data of Ref.~\cite{Walle2011} corresponding to a coverage of 0.1~ML. 
The experimental spectrum is closely reproduced by our calculation in Fig.~\ref{fig.2}(b)
but not by the calculation in Fig.~\ref{fig.2}(a). 
This observation leads us to support the proposal of Ref.~\cite{Walle2011} 
of predominantly dissociative adsorption over molecular adsorption at low coverage 
in their experiments.

\begin{figure}[t!]
\includegraphics{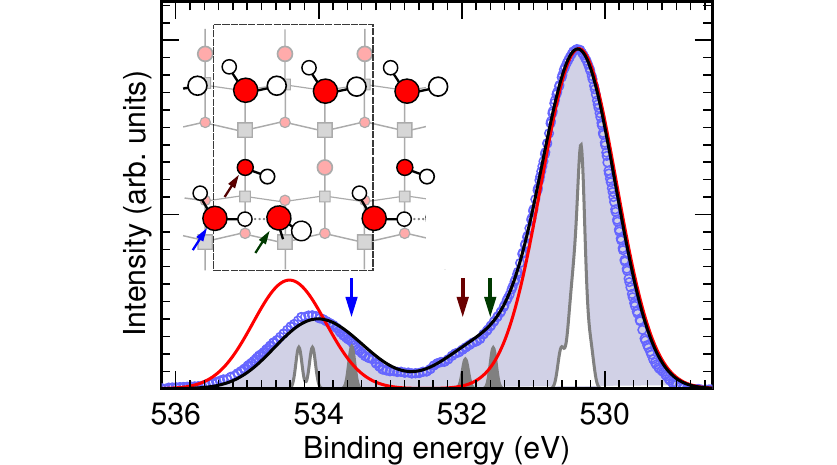}
\caption{
\label{fig.3} (Color online)
Calculated O-$1s$ core-level spectra for the TiO$_2$:H$_2$O interface at monolayer coverage, 
for molecular (red curve) and partial dissociative adsorption 
(black and gray curves), with the experimental O-$1s$ spectrum (blue circles, spectrum at 230~K 
in Fig.~3 of Ref.~\cite{Walle2011}).
The inset shows a ball-and-stick model of the partially dissociated monolayer viewed from above. 
The arrows and dark shaded area identify the O~$1s$ signature of the 
HO$\cdot\cdot\cdot$H$_2$O complex.
}
\end{figure}

Having addressed the clean surface and water at low coverage, we now consider the case 
of a full water monolayer. Following the proposal of Ref.~\cite{Walle2011} we investigated 
two interface structures: one comprising only of water molecules, and another one where 
25\% of the molecules underwent dissociation. Figure~\ref{fig.3} shows
the lowest energy structure among the four inequivalent models
allowed by our choice of surface supercell.

The spectrum calculated for the partially dissociated monolayer (black line in Fig.~\ref{fig.3})
exhibits two key differences with respect to the undissociated monolayer (red line). The first 
is a shoulder in the substrate peak around 532~eV. This effect arises from the presence of 
hydroxyl groups, as discussed in relation to Fig.~\ref{fig.2}. The second is a broadening 
and a redshift of the water peak around 534~eV. This feature is explained as follows. The 
OH group obtained from water dissociation interacts strongly with a neighboring H$_2$O molecule. 
This leads to the formation of a HO$\cdot\cdot\cdot$H$_2$O complex carrying a distinct spectral 
fingerprint (gray shaded curve in Fig.~\ref{fig.3}). In fact, the hydrogen bond between OH and H$_2$O 
(length 1.71~\AA) reduces the O-$1s$ binding energy on the coordinated 
H$_2$O molecule by 0.64~eV compared to the other water molecules. The reduction of core-level binding 
energies by H-bonds is a well-documented effect~\cite{Patrick2011,GarciaGil2013}. The broadening 
and redshift observed in Fig.~\ref{fig.3} results therefore from the presence of two different 
species on the surface, molecular H$_2$O and the HO$\cdot\cdot\cdot$H$_2$O complex.
 
When we compare our calculations with the XPS data of Ref.~\cite{Walle2011} at monolayer coverage
(Fig.~\ref{fig.3}), we find that the spectrum calculated for the 
partially dissociated water monolayer is in excellent agreement with experiment, in
contrast to that calculated for undissociated water.
This comparison supports the proposal of Ref.~\cite{Walle2011}
of dissociative water adsorption on the anatase TiO$_2$ (101) surface
in their experiments.

This result appears to contradict the commonly accepted molecular model 
of the TiO$_2$:H$_2$O interface~\cite{Vittadini1998,Sun2010,Herman2003,He2009}.
While we cannot rule out the possibility that water dissociation in the experiments 
of Ref.~\cite{Walle2011} may be induced by the X-ray irradiation~\cite{Andersson2004}, 
our present finding requires us to critically re-examine previous evidence for molecular adsorption.
In this spirit we first consider the calculation of the energetics of water adsorption.

Careful test calculations indicated that the adsorption energy 
is rather sensitive to the thickness of the TiO$_2$ model slab and the sampling of the surface 
Brillouin zone (Appendix~\ref{app.conv}). 
In the case of the undissociated water monolayer our converged adsorption energy 
(obtained using 8 TiO$_2$ layers and a 2$\times$2 surface Brillouin zone mesh) is 0.64~eV 
per water molecule. At the same time we found that the energy required for dissociating one 
in four water molecules, so as to obtain the model interface shown in Fig.~\ref{fig.3}, 
is only 60~meV.
This surprisingly small energy penalty results from the balance between 
the cost of splitting H$_2$O and the energy gained by forming the H-bond in the 
HO$\cdot\cdot\cdot$H$_2$O complex discussed above.
This is similar to the case of water adsorption on rutile TiO$_2$, where computational
and experimental studies provided evidence for such complexes~\cite{Matthiesen2009,Lindan2005}.
We found a similar trend for the case of an isolated water dimer. In this latter case we calculated
the energy required to form the HO$\cdot\cdot\cdot$H$_2$O complex to be 150~meV, i.e.\ at very
low coverage complex-assisted dissociation is only slightly less favorable than at full coverage.

Given that the calculated total energies of the undissociated and of the partially dissociated
monolayers differ by only 60~meV, we proceed with an analysis of quantum nuclear zero-point 
effects. For this purpose we investigated the vibrational eigenmodes for two model surfaces, 
with H$_2$O adsorbed either molecularly or dissociatively. We found that
upon dissociation the zero-point energy decreases by 80~meV.
This decrease can be understood by noting that the contribution to the vibrational
energy from the bending mode of the H$_2$O molecule at 1580~cm$^{-1}$ is removed upon dissociation.
After taking this zero-point correction into account, the partially (25\%) dissociated water 
monolayer becomes marginally more stable (20~meV) than the molecular monolayer.
Similarly at low coverage the formation of the H-complex is only marginally unstable
(by 70~meV) and could be temperature-activated.
The above suggests that vibrational spectroscopy could be employed to quantify the fraction of 
dissociated water, through the relative intensities of the bending mode 
at 1580~cm$^{-1}$ (present only for molecular H$_2$O) and the stretching modes at 
3700~cm$^{-1}$ (present for both molecular and dissociated H$_2$O).

Obviously the energetic differences here are so small that these results may be sensitive to 
the choice of the exchange and correlation functional, and may be affected by the neglect 
of dispersion forces, subsurface defects, and step edges~\cite{Gillan2012,Ren2012,Aschauer2010,Gong2006}.
To demonstrate this aspect we repeated full structural relaxations of the interface models
using the PBE0 hybrid functional~\cite{Adamo1999} which previously was shown to improve 
the accuracy of calculations of the vibrational
properties of liquid water~\cite{Zhang20112}.
Interestingly the formation energy of the HO$\cdot\cdot\cdot$H$_2$O complex 
starting from two water molecules is reduced when using the PBE0 hybrid functional 
(Appendix~\ref{app.pbe0}),
further demonstrating that this complex plays an important role in the structure of water on TiO$_2$.

Our present analysis of the adsorption energetics indicates that there may not necessarily
be a contradiction between the 
spectroscopic observation of dissociated water and DFT calculations of adsorption energies. 
Previous calculations taken as evidence against dissociative adsorption, although technically correct, 
did not consider the possibility of forming the energetically favorable HO$\cdot\cdot\cdot$H$_2$O 
complex, thus reaching conclusions that water adsorbs molecularly.  Our calculations 
indicate that complex-assisted dissociation is slightly more favorable than molecular adsorption.

\begin{figure}[t!]
\includegraphics[width=7.5cm]{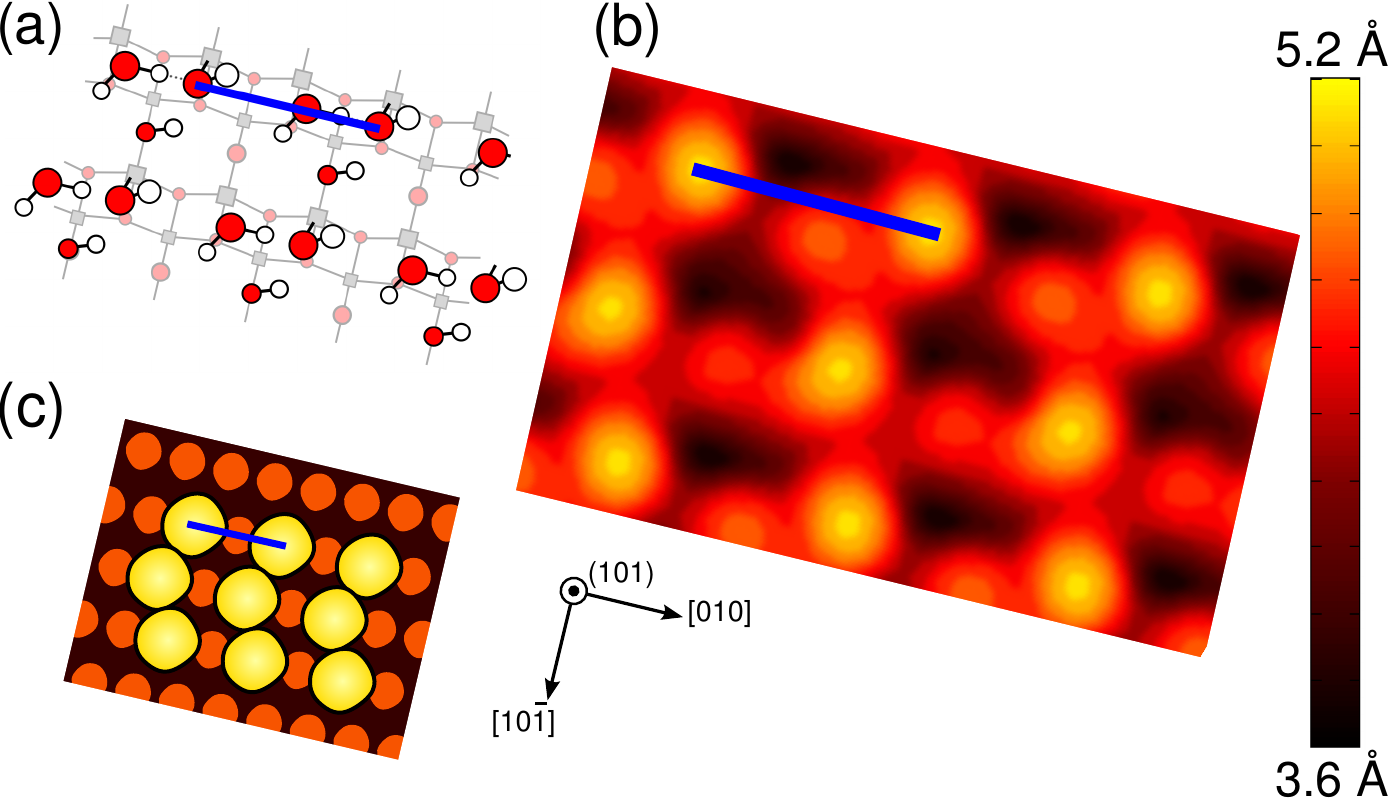}
\caption{
\label{fig.4} (Color online)
Ball-and-stick representation (a) and calculated STM map (b) for a monolayer of HO$\cdot\cdot\cdot$H$_2$O 
complexes. A cartoon sketch to scale of the images reported in Ref.~\cite{He2009} featuring the 
experimentally-measured topography is shown in (c). 
The STM map was calculated~\cite{Tersoff1985} as an isovalue 
plot ($3.0\times10^{-4}$~electrons~\AA$^{-3}$) of the local density-of-states integrated over an energy 
window of 2.1~eV above the conduction band edge of TiO$_2$. The blue segment is along the [010] direction 
and measures 7.6~\AA. The elevation is given with respect to the fivefold coordinated Ti sublattice.
}
\end{figure}

Having identified that the HO$\cdot\cdot\cdot$H$_2$O complex might play an important 
role in the  adsorption of water on anatase TiO$_2$ (101), 
we wanted to check whether this complex might
be related to the 2$\times$2 superstructure observed in the STM experiments 
of Ref.~\cite{He2009} at high coverage. To this aim we calculated the STM map of a monolayer 
of the HO$\cdot\cdot\cdot$H$_2$O complex in the Tersoff-Hamann approximation~\cite{Tersoff1985} 
(Fig.~\ref{fig.4}). Our choice of performing the calculation for a full monolayer is dictated
by computational convenience, and does not affect our analysis since the STM map is effectively
a local density of states \cite{Tersoff1985}.
 The calculated STM map consists of bright spots aligned along the [010] 
direction, derived from the empty O~2$p$ states of the OH groups.
Interestingly we found that our calculated maps are in agreement with the topography 
measured in Ref.~\cite{He2009}, accounting for the doubling of the period 
along the [010] direction.
Reproducing this period-doubling with exclusively molecular water requires
that H$_2$O molecules do not adsorb on adjacent sites~\cite{He2009} (c.f.\ Appendix~\ref{app.stm}).
A partially-dissociated monolayer with a significant fraction of 
HO$\cdot\cdot\cdot$H$_2$O complexes provides an alternative explanation 
of the measured STM images.
Additional discussion of experimental studies of the
anatase TiO$_2$:H$_2$O interface is provided in Appendix~\ref{app.discussion}.

\section{Conclusions}

In summary, our first-principles calculations support the assignment
of dissociated water in the XPS
experiments of Ref.~\cite{Walle2011} on the TiO$_2$ anatase (101) surface
at monolayer coverage. In this configuration water dissociation is enabled by the H-bonded 
complex HO$\cdot\cdot\cdot$H$_2$O. We demonstrate that this complex is energetically
more favorable than molecular adsorption, and that it can provide an alternative explanation
for the 2$\times$2 superstructure observed in the STM images of Ref.~\cite{He2009}.
Taken together our present findings outline a structural model 
of the TiO$_2$:H$_2$O interface
which is compatible with a variety of experimental and theoretical studies 
on this important photocatalytic interface. 
More generally this work demonstrates how the reverse-engineering of 
experimental data using {\it ab initio} spectroscopy represents a powerful tool for
elucidating the nature of complex materials for energy applications.

\begin{acknowledgments}
This work was supported by the European Research Council (EU FP7 / ERC grant no. 239578), 
the UK Engineering and Physical Sciences Research Council (Grant No. EP/J009857/1) and 
the Leverhulme Trust (Grant RL-2012-001). Calculations were performed at the Oxford Supercomputing 
Centre and at the Oxford Materials Modelling Laboratory. Figures rendered using VMD~\cite{VMD}.
\end{acknowledgments}

\appendix

\section{Computational details}

\subsection{DFT calculations}
\label{app.dft}
The planewave kinetic energy cutoffs of wavefunctions and charge density were 35 and 200~Ry,
respectively. The core-valence interaction was treated by means of ultrasoft
pseudopotentials~\cite{Vanderbilt1990}, with the semicore Ti~3$s$ and 3$p$ states explicitly
described.  The TiO$_2$ substrate was modeled as a stoichiometric slab, obtained
by cutting bulk anatase along the (101) plane~\cite{Lazzeri2001,Patrick2011}.
The resulting slab consists of 8 TiO$_2$ layers (12.9~\AA \ thickness), with the periodic
slab replicas separated by a vacuum region of 15~\AA. The minimal surface cell considered here
has dimensions 10.4$\times$7.6~\AA$^2$ and contains 4~TiO$_2$ units. For the calculation
of interface geometries, adsorption energies and STM maps we sampled the surface Brillouin zone
using a 2$\times$2 grid, while for calculations of core-level shifts we used a 2$\times$2
supercell (monolayer coverage) and a 1$\times$2 supercell (low coverage), both with
$\Gamma$-point sampling. Calculations using
the 2$\times$2 surface supercell (432 atoms) 
were necessary in order to minimize the 
interaction of charged replicas~\cite{Binder2012,Makov1995}.
Geometry optimizations were performed with the bottom TiO$_2$ layer fixed in order to mimic
a semi-infinite substrate, and carried out until the forces on all the other atoms were below
30~meV/\AA.  The vibrational eigenmodes and eigenfrequencies for the calculation of zero-point
energies were obtained within the harmonic approximation, using the method of finite
displacements~\cite{Ackland1997} and the minimal surface unit cell.

\subsection{Sensitivity to electron escape depth}
\label{app.lambda}

\begin{figure*}
\includegraphics[width=120mm]{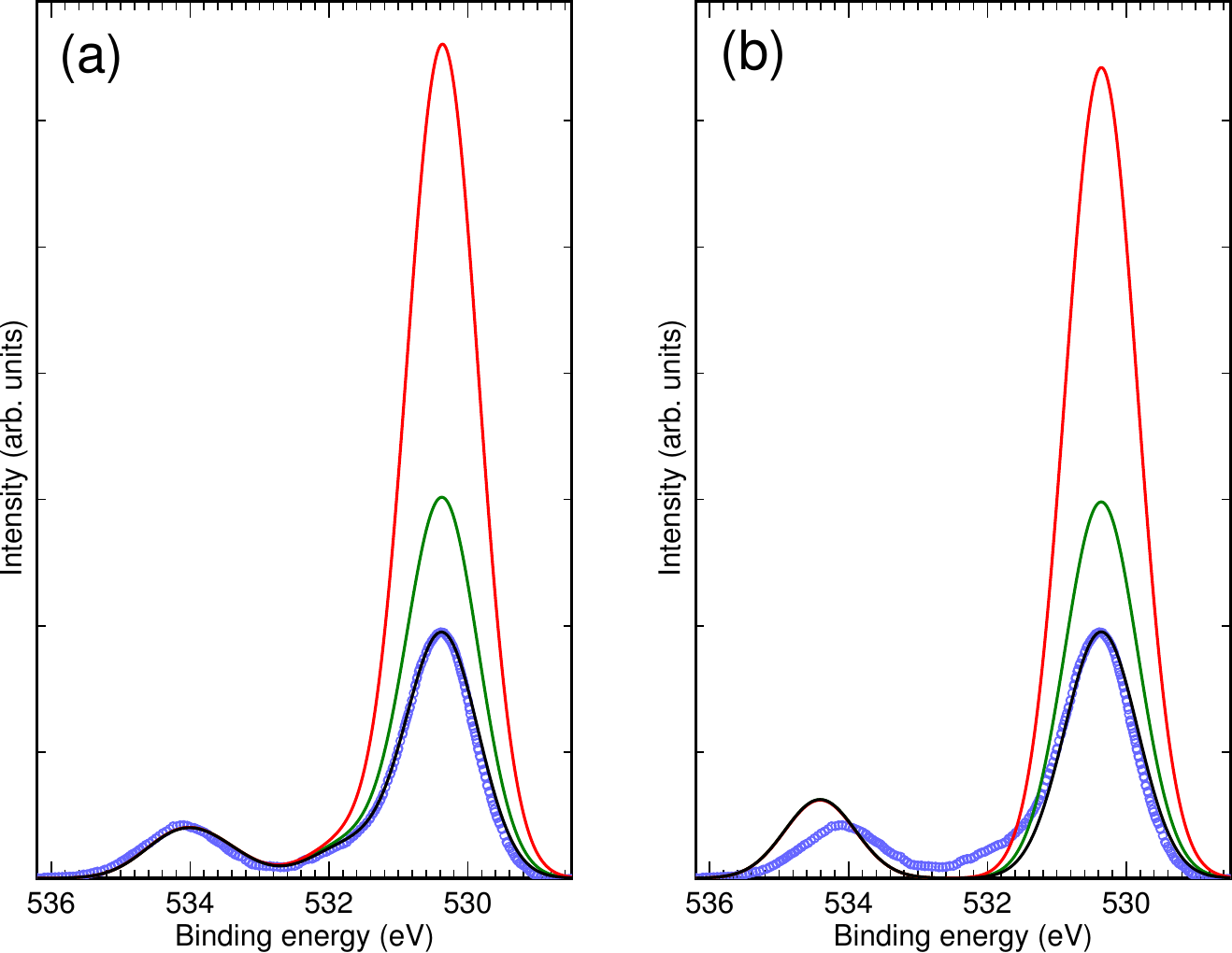}
\caption{
Calculated O-$1s$ core-level spectra for the TiO$_2$:H$_2$O interface at monolayer coverage 
in the case of 25\% H$_2$O dissociation (a) and fully molecular adsorption (b) for
different values of the electron escape depth $\lambda$.
The black, green and red lines correspond to $\lambda =$ 3.5, 5.0 and 10.0~\AA \ respectively.
The experimental O-$1s$ spectrum (blue circles, spectrum at 230~K
in Fig.~3 of Ref.~\cite{Walle2011}) is also shown.
\label{fig.app1}
}
\end{figure*}

Throughout our study we used a value of 3.5~\AA \ for the electron escape
depth $\lambda$, which is physically reasonable based on the energies
of the incident X-rays and electron binding energies~\cite{Hufnerbook2}.
In Fig.~\ref{fig.app1} we illustrate
the effect of using different values for $\lambda$.
It can be seen that the escape depth essentially determines
the relative strength of the bulk and adsorbate contributions
to the spectrum.
Intuitively, larger values of $\lambda$ reduce the surface
sensitivity.
We find that the structure of the adsorbate contribution
at 532~eV and above
is insensitive to $\lambda$.
This observation can be rationalized by noting that the oxygen
atoms of molecular and dissociated H$_2$O are located at similar
distances from the surface.
Comparing to the experimental data of Ref.~\cite{Walle2011} in
Fig.~\ref{fig.app1}(b)
shows how choosing different values of $\lambda$ cannot remedy the failure
of the purely molecular monolayer to account for the
experimentally-observed features.

\subsection{Dissociative adsorption of H$_2$O at low coverage}
\label{app.OH}
\begin{figure}
\includegraphics{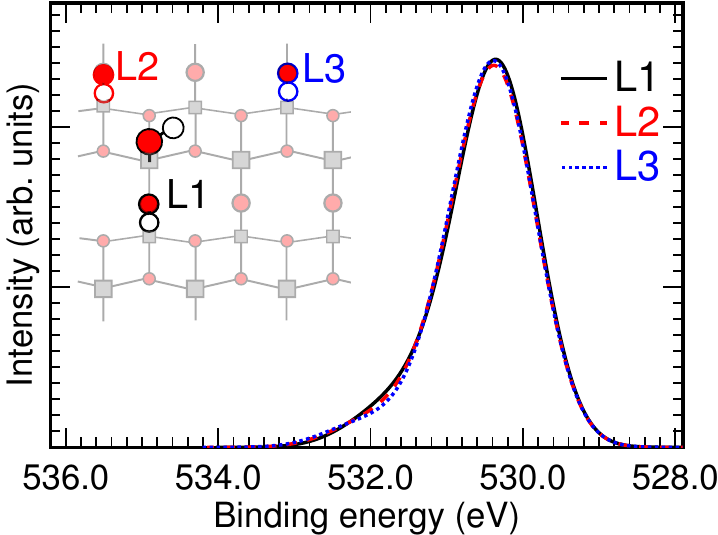}
\caption{
Calculated O-$1s$ core-level spectra for the TiO$_2$:H$_2$O interface at low coverage in the case
of dissociative adsorption, with the
OH groups located at the different sites marked L1, L2, and L3.
The spectra (shown as solid, red dashed, and blue dotted curves respectively) were calculated
with a gaussian broadening of 1.2~eV, and are practically indistinguishable.
\label{fig.app2}
}
\end{figure}

In dissociative adsorption of H$_2$O the OH group binds covalently 
to a fivefold Ti atom (bond length 1.82~\AA), and the remaining proton binds to a twofold O atom,
thereby yielding an additional hydroxyl group.
For completeness we explored 
models differing in the separation between these two OH groups (Fig.~\ref{fig.app2}).
The model with the additional hydroxyl group at the site labeled by L1 in Fig.~\ref{fig.app2} 
has previously been denoted the ``intrapair configuration''~\cite{Vittadini1998} or ``pseudo-dissociated 
water''~\cite{Walle2011}.
The model with the additional OH group at site L2 was denoted the ``interpair 
configuration''~\cite{Vittadini1998}.
In accordance with the findings of Ref.~\cite{Vittadini1998}, we found the intrapair
configuration (L1) to be more stable than interpair (L2), by 0.05~eV per H$_2$O molecule.
Configuration L3 is taken to describe the limit of isolated 
OH groups.
Note that the description of L3 requires the use of a 1$\times$2 supercell.

When calculating the  O-1$s$ spectrum for models L1--L3, in each case there are two
new contributions to the spectrum from the inequivalent OH groups.
As discussed in the main text, for L1 these contributions appear at 531.34 and 531.93~eV.
Similarly, for configurations L2 and L3 the binding energies of the corresponding 
O atoms are 531.21/532.05~eV and 531.06/532.11~eV, respectively.
These small differences are completely washed out when vibrational broadening is included. 
In fact Fig.~\ref{fig.app2}
shows that the spectra of the various configurations of dissociated water at low coverage 
are practically indistinguishable. This finding indicates that hydroxyl-hydroxyl interactions 
can safely be neglected in the interpretation of O-1$s$ core-level spectra.

\subsection{Numerical convergence of adsorption energies}
\label{app.conv}

Since the presence of dissociated water on the anatase TiO$_2$ (101) surface
is at variance with the commonly accepted molecular model of the
TiO$_2$:H$_2$O interface~\cite{Vittadini1998,Sun2010,Herman2003,He2009}, we re-examined the 
adsorption energetics.
In this section we focus on numerical convergence.

\begin{figure*}
\includegraphics[height=120mm]{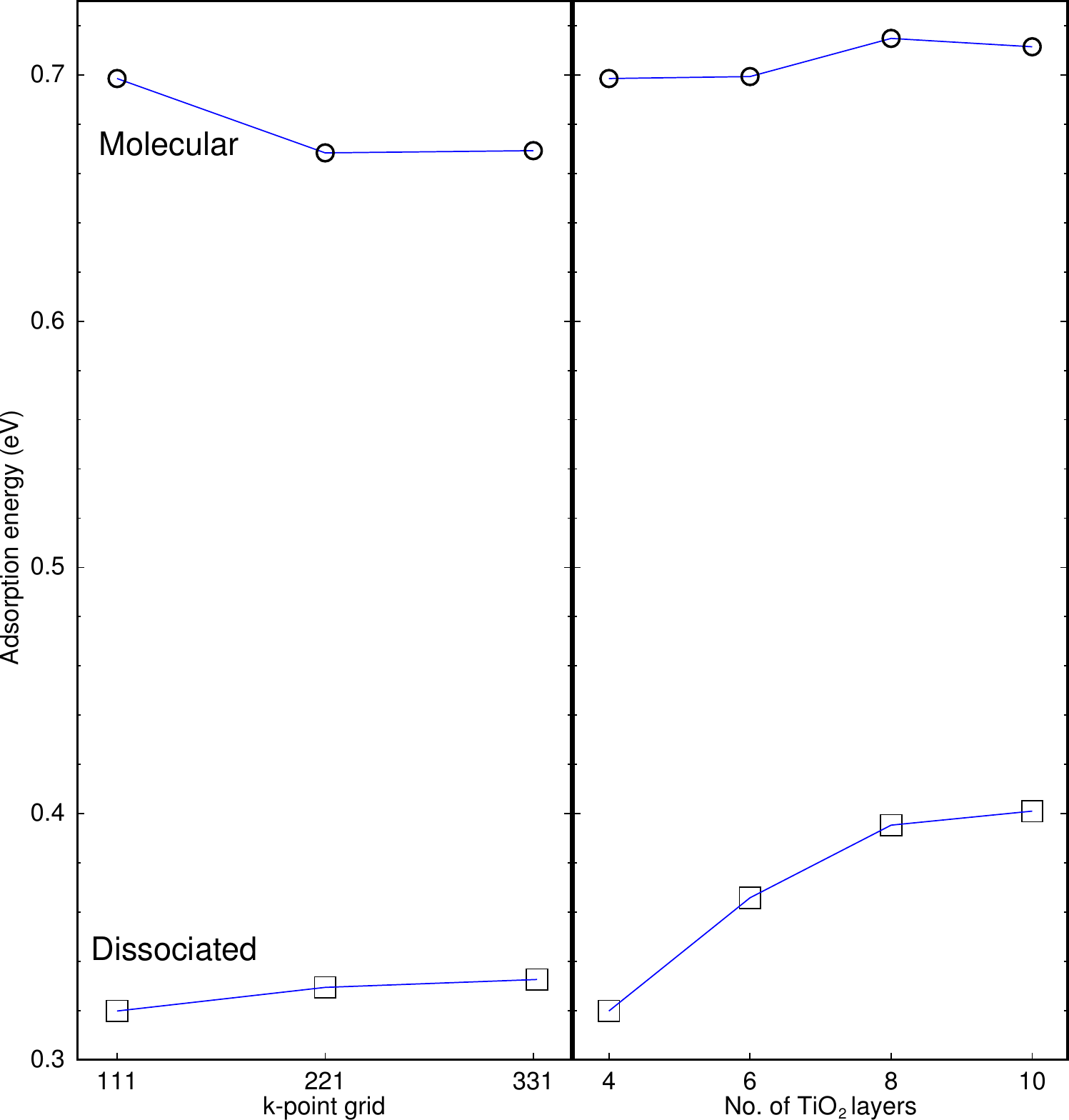}
\caption{
Convergence of the adsorption energies for molecular (circles) and dissociative  (squares) 
H$_2$O adsorption as a function of $k$-point sampling and thickness of
the TiO$_2$ slab. 
For dissociative adsorption the intrapair configuration
(L1 in Fig.~\ref{fig.app2}) was considered.
In the test of convergence with $k$-points a slab of 
4 TiO$_2$ layers was used, while in the test of convergence with slab thickness
$\Gamma$-point sampling was used.  
All other computational details were  
as reported in Appendix~\ref{app.dft}.
The blue lines are guides to the eye.
\label{fig.app3}
}
\end{figure*}

Figure~\ref{fig.app3} displays the adsorption energies of isolated
molecular and dissociated H$_2$O on the anatase TiO$_2$ (101) surface, calculated as a
function of the surface Brillouin zone ($k$-point) sampling or TiO$_2$ slab thickness.
Using $\Gamma$-point sampling and a slab of 4 TiO$_2$ layers (the same
computational setup as Ref.~\cite{Vittadini1998}) we calculate adsorption
energies of 0.70~and~0.32~eV for molecular and dissociated water, respectively (difference 0.38~eV).
Keeping the slab thickness fixed at 4 TiO$_2$ layers but increasing the Brillouin zone sampling to 2$\times$2 (left
plot of Fig.~\ref{fig.app3}) causes small changes in the adsorption energies, to 0.67~and~0.33~eV
(difference 0.34~eV);
further increasing the sampling to 3$\times$3 results in changes of less than 0.005~eV to the
adsorption energies.
Thus we see that the difference between molecular and dissociative adsorption energies reduces
by 0.04~eV upon converging with $k$-points.

Now fixing the Brillouin zone sampling to the $\Gamma$-point and instead 
varying the slab thickness (right plot of Fig.~\ref{fig.app3}) we observe
differing convergence behavior for molecular and dissociative adsorption.
Upon increasing the slab thickness through the sequence 4$\rightarrow$6$\rightarrow$8$\rightarrow$10
layers, the molecular adsorption energy takes the values
0.70$\rightarrow$0.70$\rightarrow$0.71$\rightarrow$0.71~eV, i.e.\ a change of 0.01~eV
on tripling the thickness of the slab (5.8--16.4~\AA).
Meanwhile the variation in adsorption energy for dissociated H$_2$O is an order
of magnitude larger, taking the values
0.32$\rightarrow$0.37$\rightarrow$0.40$\rightarrow$0.40~eV.
Thus the difference between molecular and dissociative adsorption energies decreases
from 0.38~eV to 0.31~eV after converging the results with the thickness of the TiO$_2$ slab.

Our convergence tests reveal that
increasing either $k$-point sampling or slab thickness
reduces the difference in molecular and dissociative adsorption energies.
Converging both quantities together further reduces
this difference.
Using a 2$\times$2 $k$-point
sampling and an 8-layer TiO$_2$ slab we obtain adsorption energies of 0.69~eV and 0.42~eV
for molecular and dissociated H$_2$O.
The obtained difference in adsorption energies, 0.27~eV, is 0.11~eV smaller than
the unconverged case.
As a footnote, we note that our calculations highlight a potential danger with
using only the molecular adsorption energy to test convergence, 
since this quantity changes by less than 0.001~eV on moving from a 4 to a 6-layer slab.

\subsection{Hybrid functional calculations of adsorption energetics}
\label{app.pbe0}

Our work has shown that careful numerical convergence, combined with consideration of
quantum nuclear zero-point effects, leads to energetic differences between molecular
and partially dissociated models of the  TiO$_2$:H$_2$O interface which are
very small; at monolayer coverage, the partially dissociated monolayer is
favorable by 20~meV, while at low coverage the cost of forming the 
H-bonded HO$\cdot\cdot\cdot$H$_2$O complex was found to be 70~meV.
These energy differences are so small that the results may be sensitive to
the choice of the exchange and correlation functional.
To explore this point further we decided to repeat our calculations of
adsorption geometries and energies using the PBE0 hybrid functional~\cite{Adamo1999},
which has previously been shown to improve the accuracy of calculations of
the vibrational properties of liquid water~\cite{Zhang20112}.

A difficulty with hybrid functional approaches, particularly when combined with
planewave basis sets, is their significant computational expense as compared
to local or semilocal functionals.
Furthermore, the implementation of hybrid functional calculations in the current general release 
of \texttt{Quantum ESPRESSO} is not compatible with ultrasoft pseudopotentials.
Therefore, we employed a reduced computational setup for the hybrid functional calculations,
and used norm-conserving pseudopotentials.
In order that we might assess the effect of the exact
exchange directly, 
we also performed additional calculations with an identical computational setup
at the PBE level of theory.
Finally we compare the values calculated within the reduced computational setup 
with the fully converged (conv.) PBE values, calculated as described
in Appendix~\ref{app.dft}.

The reduced computational setup is described as follows.
The core-valence interaction was treated by means of norm-conserving
pseudopotentials, with the Ti~3$s$ and 3$p$ states frozen into the core.
A converged value of 100~Ry was used for the planewave kinetic 
energy cutoff of the wavefunctions.
A 1$\times$1 surface cell was used to construct structural models, as for the fully-converged
setup;
however the slab thickness was reduced to 4 TiO$_2$ layers and the Brillouin zone
sampled at the $\Gamma$-point only.
The lattice parameters used to construct the slab were obtained
by relaxing anatase TiO$_2$ in bulk with the norm-conserving
pseudopotentials at the PBE level of theory, sampling the Brillouin
zone on a 6$\times$6$\times$6 mesh.
The periodic slab replicas were separated by a vacuum region of 9~\AA.
Full geometry optimizations were performed exactly as in the fully-converged setup, with the bottom 
TiO$_2$ layer fixed in order to mimic
a semi-infinite layer and carried out until the forces on all the other atoms were below
30~meV/\AA.

\begin{figure*}
\includegraphics[width=150mm]{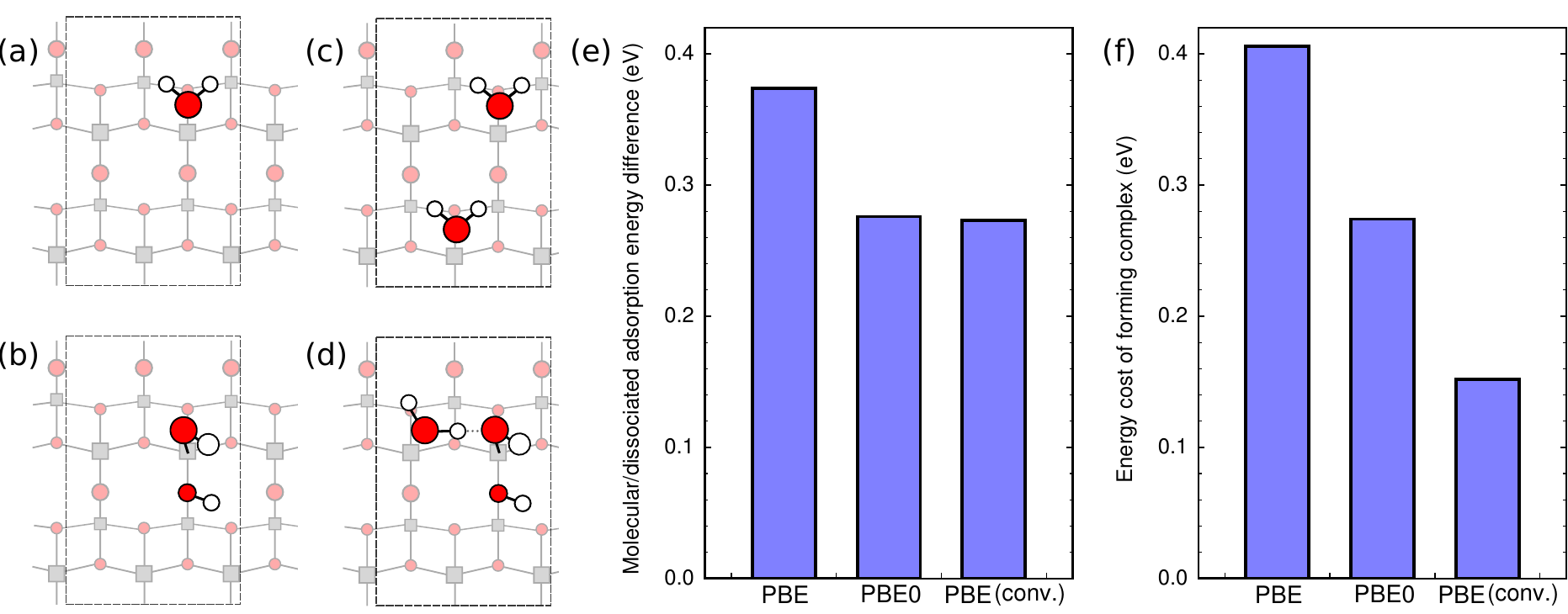}
\caption{
(a)--(d): Ball-and-stick representations of H$_2$O adsorption models
used to test the effect of including exact exchange into calculations
of adsorption energies, corresponding to (a) molecular adsorption, (b) dissociated
adsorption, (c) two molecularly-adsorbed H$_2$O molecules per unit
cell and (d) the H-bonded HO$\cdot\cdot\cdot$H$_2$O complex.
The dashed-line rectangles show the surface unit cell.
(e), (f): Bar charts giving adsorption energies calculated using
different approaches.
The labels ``PBE'' and ``PBE0'' denote the exchange-correlation functionals
used to calculate adsorption energies within the reduced computational 
setup.
``PBE (conv.)'' refers to calculations performed with the converged computational setup
described in Appendix~\ref{app.dft} and used throughout
our study.
The bars shown in (e) give the difference in molecular and dissociative
adsorption energies, obtained as the total energy difference between
the models shown in (a) and (b).
The bars shown in (f) give the energy cost of forming the HO$\cdot\cdot\cdot$H$_2$O complex, 
obtained as the total energy difference between
the models shown in (c) and (d).
\label{fig.app4}
}
\end{figure*}

The structural models considered at the PBE0 level of theory are shown
in Fig.~\ref{fig.app4}.
We obtained the fully-relaxed geometries of (a) molecularly-adsorbed
H$_2$O at low coverage, (b) dissociated (intrapair) H$_2$O at low coverage,
(c) two molecularly-adsorbed H$_2$O molecules and (d) the H-bonded HO$\cdot\cdot\cdot$H$_2$O complex.
We also obtained the structure of the clean anatase TiO$_2$ (101) slab and of the isolated
H$_2$O molecule.

\begin{table}
\small
\begin{tabular}{l c c c c}
\hline
           & Molecular& Dissociated & Difference & Complex \\ 
\hline
PBE        & 0.80     & 0.42        & 0.38       & 0.41 \\
PBE0       & 0.84     & 0.56        & 0.28       & 0.27 \\
PBE (conv.)& 0.69     & 0.42        & 0.27       & 0.15 \\
\hline
\end{tabular}
\caption{
\label{tab.app1}
Adsorption energetics calculated with the reduced computational setup
at the PBE or PBE0 level of theory, or at the PBE level of theory at the
converged (conv.) computational setup.
The corresponding models of molecular and dissociated
H$_2$O on the anatase TiO$_2$ (101) surface are shown in Figs.~\ref{fig.app4}(a)~and~(b).
The table also gives the difference in molecular and dissociated adsorption energies,
and the energy cost of forming the H-bonded HO$\cdot\cdot\cdot$H$_2$O complex
from two molecularly-adsorbed H$_2$O molecules [Figs.~\ref{fig.app4}(c)~and~(d)].
The data of the last two columns are displayed as bar charts in Figs.~\ref{fig.app4}(e)~and~(f).
All values are in eV.
}
\end{table}

Table~\ref{tab.app1} gives the adsorption energies calculated with the reduced 
computational setup using either the PBE or PBE0 functional to account 
for exchange and correlation.
For comparison we also provide the same quantities calculated with the PBE
functional and the fully-converged (conv.) computational setup.
The table shows the adsorption energies calculated for molecular and dissociated water,
and their difference, and also gives the energy cost of forming the HO$\cdot\cdot\cdot$H$_2$O complex.
The latter is obtained as a total energy difference between the models shown in Figs.~\ref{fig.app4}(c)~and~(d).
The data of the last two columns of Table~\ref{tab.app1} are displayed as bar charts in Figs.~\ref{fig.app4}(e)~and~(f).

Our hybrid functional calculations have produced some interesting results.
First, comparing the energies for molecular and dissociative adsorption at
the reduced computational setup, we find that the PBE0 hybrid functional 
predicts an increased stability of dissociative adsorption compared to PBE.
Overall the difference
in molecular and dissociative adsorption energies decreases by 0.10~eV in the PBE0 calculations
[Fig.~\ref{fig.app4}(e)].
Further comparing the PBE calculations obtained for the reduced and converged computational
setups we find this same difference to decrease by 0.11~eV.
Therefore we expect that
the calculated difference between molecular
and dissociative H$_2$O adsorption with PBE0 of 0.28~eV
represents an upper bound, and further converging these
calculations with $k$-points and slab thickness
would further reduce this difference.

Second, considering the energy cost of forming the complex [Fig.~\ref{fig.app4}(f)],
we again find a reduction on moving to the PBE0 level of theory, by 0.14~eV.
This reduction can be attributed to the lower cost of water dissociation
in PBE0 (0.10~eV, as discussed above)
and to a stronger hydrogen bond in the 
HO$\cdot\cdot\cdot$H$_2$O complex (0.04~eV). 
Again noting that the energy cost of forming the complex decreases by 0.26~eV
on moving from the
reduced to the converged computational setups at the PBE level of theory, we are left with a possibility 
that a fully-converged PBE0 calculation would predict the HO$\cdot\cdot\cdot$H$_2$O
complex to be stable even before considering the vibrational contribution to the energy.

As well as adsorption energetics, the full geometry relaxations
provide insight into the different descriptions of the structure of the 
H-bonded HO$\cdot\cdot\cdot$H$_2$O complex in PBE and PBE0.
We found that the length of the H-bond [dotted line in Fig.~\ref{fig.app4}(d)] 
increases by just 0.012~\AA \ on moving from PBE to PBE0.
This observation can be explained simply by noting that the formation
of the HO$\cdot\cdot\cdot$H$_2$O complex is determined by the competition
of the covalent Ti--O bond of the hydroxyl, the covalent Ti--O bond of
the H$_2$O molecule, and the H-bond between the hydroxyl and H$_2$O molecule.
Given the enhanced strength of covalent bonds compared to H-bonds,
it is not surprising that there are fewer changes observed on moving
from a semilocal to nonlocal description of exchange compared to 
that observed for homogeneous molecular water~\cite{Zhang20112}, where H-bonds dominate.

Clearly, in light of the discussion of numerical convergence
in the previous section, the values presented in Table~\ref{tab.app1} and Figs.~\ref{fig.app4}(e)~and~(f)
cannot be considered definitive.
However, the hybrid functional calculations clearly show  that the energetic
balance between molecular and dissociative H$_2$O adsorption on the anatase TiO$_2$ (101)
surface is extremely delicate.
As a consequence, we conclude that current first-principles calculations
cannot rule out the possibility of water dissociation on this surface.

\subsection{STM image of the molecular monolayer}
\label{app.stm}

\begin{figure}
\includegraphics[width=7.5cm]{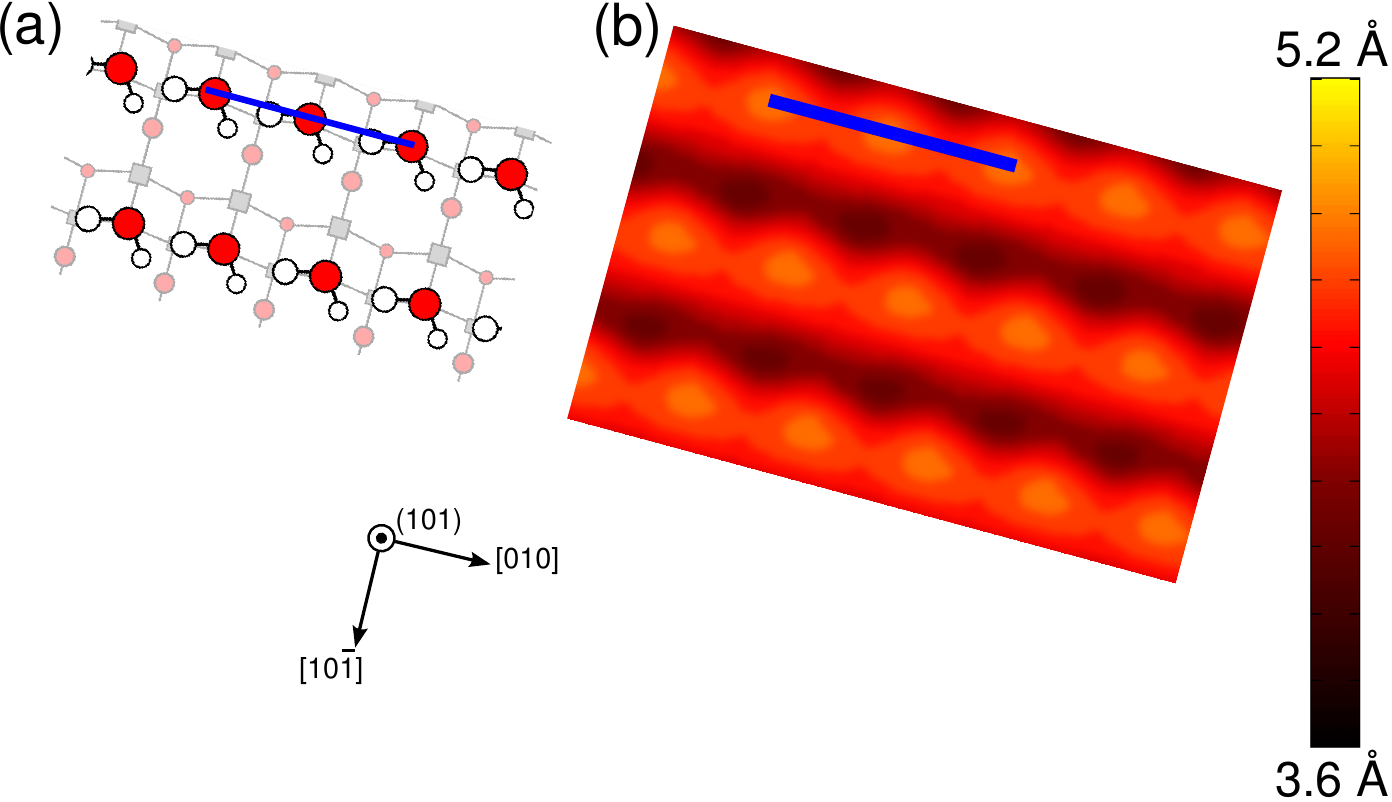}
\caption{
Ball-and-stick model (a) and calculated STM image (b) of
purely molecularly adsorbed water.
The STM image was obtained using the same computational
setup as in Fig.~4 of the main text.
As in Fig.~4 the blue segment measures 7.6~\AA. 
\label{fig.app5}
}
\end{figure}

Figure~\ref{fig.app5} shows the calculated
STM image of purely molecularly-adsorbed water.
Each adsorbed water molecule appears as a bright
spot in the spectrum, with the calculated image
thus showing a 1$\times$1 periodicity.
In order to account for the experimentally-observed 
2$\times$2 periodicity using only molecular water,
it is necessary for water not to adsorb at adjacent
sites but rather be staggered, as proposed in Ref.~\cite{He2009}.

\section{Experimental studies of water adsorption at the anatase TiO$_2$:H$_2$O interface}
\label{app.discussion}

In 1998, the authors of Ref.~\cite{Vittadini1998} considered
previous experimental studies of the anatase TiO$_2$:H$_2$O interface
and noted that
``On the basis of this experimental information, different models of the anatase surface and of water
adsorption at this surface have been proposed, but no clear picture has emerged yet.''
Subsequently, to our knowledge three experimental works were published aimed at
obtaining a clearer picture of the behavior of water on well-defined
anatase (101) surfaces: a combined temperature-programmed
desorption and XPS study~\cite{Herman2003}, an STM study~\cite{He2009}
and another XPS study published more recently~\cite{Walle2011}.

Our current work supports the interpretation of the XPS 
measurements of Ref.~\cite{Walle2011}
as evidence for water dissociation.
We also showed that features found in the STM images measured 
in Ref.~\cite{He2009} (the 2$\times$2 superstructure and 
zigzag pattern) can also be explained in terms of a partially dissociated monolayer
of H$_2$O, although the authors of Ref.~\cite{He2009} 
provided an alternative explanation in terms of a purely molecular
monolayer.

The remaining experimental work is the TPD/XPS study of Ref.~\cite{Herman2003}.
First considering the XPS, comparing the measurements of Ref.~\cite{Herman2003} 
to Ref.~\cite{Walle2011}
shows that the latter has achieved the higher resolution in measuring the O-$1s$ spectrum.
This higher resolution is essential to discern
the presence of OH groups, since they appear only as a shoulder to the 
bulk TiO$_2$ peak [c.f.\ Fig.~2(b) of our work].
Partial dissociation of H$_2$O stabilized by H-bonding interactions also
provides an explanation for the coverage-dependent
shift of the molecular H$_2$O peak observed in 
Ref.~\cite{Herman2003}, 
since our calculations showed that this feature is sensitive to
 intermolecular H-bonding.

Now turning to the TPD measurements, the TPD spectrum of hydrated anatase
shows a broad peak at 250K, and additional features at 190 and 160 K at coverages
greater than one monolayer.
In contrast to experiments performed on the rutile (110) surface~\cite{Beck1986},
there is no peak at $\sim$500~K.
The absence of this peak was considered significant, since the 500~K feature in rutile
was assigned to dissociated water in Ref.~\cite{Beck1986}.
However, as noted by the authors of Ref.~\cite{Herman2003}, the hydroxyls
found on the rutile (110) surface are thought to form at O vacancies; therefore,
it is not clear how the behavior of hydroxyl species formed at defects on rutile 
should correspond
to that of dissociated water on the stoichiometic anatase surface.
Indeed we note that the calculations of Ref.~\cite{Schaub2001}
found an adsorption
energy of 0.94~eV for dissociated H$_2$O on the defective rutile (110) surface, which
is larger than the adsorption energies calculated for water on stoichiometric anatase.
In fact, as shown by the XPS measurements  at different temperatures reported in Ref.~\cite{Walle2011},
the hydroxyl groups on the anatase (101) surface observed by the XPS have all been desorbed by 400 K.

The question remains therefore of how to assign the broad peak observed 
in the TPD spectrum at 250~K.
The authors of Ref.~\cite{Herman2003} assigned the peak
as originating from the desorption molecular water only, 
on the basis that the adsorption energy obtained
by them through a Redhead analysis was found to be similar to that obtained in
the previous theoretical work of Ref.~\cite{Vittadini1998}.
However, if the adsorption energies of molecular and dissociated H$_2$O
are similar, as showed by our work, 
the broad peak at 250~K
could equally be assigned to the desorption of both molecular and dissociated
H$_2$O.

\end{document}